\pdfoutput=1

\documentclass[onecolumn]{emulateapj}

\usepackage{amsmath}
\usepackage{graphicx}
\usepackage{color}
\usepackage[colorlinks]{hyperref}
\hypersetup{
    colorlinks,	
    citecolor=blue,
}

\newcommand{\be}{\begin{eqnarray}}
\newcommand{\ee}{\end{eqnarray}}

\def\jcap{JCAP}

\shorttitle{Testing GR using the continuum-fitting method}
\shortauthors{Tripathi et al.}

\begin{document}

\title{Testing general relativity with the stellar-mass black hole in LMC~X-1\\using the continuum-fitting method}

\author{Ashutosh~Tripathi\altaffilmark{1}, Menglei~Zhou\altaffilmark{1}, Askar~B.~Abdikamalov\altaffilmark{1}, Dimitry~Ayzenberg\altaffilmark{1}, Cosimo~Bambi\altaffilmark{1,\dag}, Lijun~Gou\altaffilmark{2,3}, Victoria~Grinberg\altaffilmark{4}, Honghui~Liu\altaffilmark{1}, and James~F.~Steiner\altaffilmark{5}}

\altaffiltext{1}{Center for Field Theory and Particle Physics and Department of Physics, 
Fudan University, 200438 Shanghai, China. \email[\dag E-mail: ]{bambi@fudan.edu.cn}}
\altaffiltext{2}{Key Laboratory for Computational Astrophysics, National Astronomical Observatories, Chinese Academy of Sciences, 100012 Beijing, China}
\altaffiltext{3}{School of Astronomy and Space Science, University of Chinese Academy of Sciences, 100049 Beijing, China}
\altaffiltext{4}{Institut f\"ur Astronomie und Astrophysik (IAAT), Eberhard-Karls Universit\"at T\"ubingen, 72076 T\"ubingen, Germany}
\altaffiltext{5}{Harvard-Smithsonian Center for Astrophysics, Cambridge, MA 02138, United States}

\begin{abstract}
The iron line and the continuum-fitting methods are currently the two leading techniques for measuring black hole spins with electromagnetic radiation. They can be naturally extended for probing the spacetime geometry around black holes and testing general relativity in the strong field regime. In the past couple of years, there has been significant work to use the iron line method to test the nature of black holes. Here we use the continuum-fitting method and we show its capability of constraining the spacetime geometry around black holes by analyzing 17~\textsl{RXTE} data of the X-ray binary LMC~X-1.
\end{abstract}

\keywords{Kerr metric --- astrophysical black holes --- X-ray astronomy}


\section{introduction}

Einstein's theory of general relativity was proposed over a century ago and has successfully passed a large number of observational tests, mainly in the weak field regime~\citep{will}. Thanks to new observational facilities, nowadays tests of general relativity in the strong field regime are of broad interest and undergoing intense study. Astrophysical black holes are ideal laboratories for testing strong gravity. The spacetime geometry around these objects is thought to be well approximated by the Kerr solution~\citep{k1,k2}, as deviations induced by a non-vanishing electric charge or by the presence of accretion disks or nearby stars are usually very small~\citep{d1,d2,d3}. On the other hand, macroscopic deviations from the Kerr geometry are predicted in models with exotic fields~\citep[e.g.][]{e1}, in a number of modified theories of gravity~\citep[e.g.][]{e2}, or as a result of large quantum gravity effects near the black hole event horizon~\citep[e.g.][]{e3,e4,e5}.

Black holes can be tested either with electromagnetic or gravitational wave techniques, and the two approaches are complementary. Electromagnetic tests~\citep{rmp,2016CQGra..33l4001J,2018GReGr..50..100K}, strictly speaking, are more suitable to probe the interplay between the gravity and the matter sectors (geodesic motion of particles and non-gravitational physics in presence of gravity). Gravitational wave tests~\citep{2013LRR....16....7G,2013LRR....16....9Y,gw1} can probe Einstein's theory of general relativity in the dynamical regime. It may be possible that new physics only shows up in one of the two spectra (i.e. either the electromagnetic or the gravitational wave spectrum) and not in the other one~\citep{will,2008PhRvL.101i9001B,2014JCAP...03..034B,2019arXiv191212629L}. When the two approaches can be compared to test the Kerr metric around black holes, they seem to provide similar constraints~\citep{gw2}.

The two leading electromagnetic techniques for measuring black hole spins under the assumption of the Kerr metric are X-ray reflection spectroscopy~\citep{i1,i2}, aka analysis of the iron line and the Compton hump, and the continuum-fitting method~\citep{cfm1,cfm2}, or analysis of the thermal spectrum. Both techniques can be naturally extended to test the Kerr black hole hypothesis~\citep{t1,t2,t3,t4,t5}. In the past few years, there has been significant work to test black holes using X-ray reflection spectroscopy and there are now a number of published results on observational constraints using \textsl{XMM-Newton}, \textsl{Suzaku}, and \textsl{NuSTAR} data~\citep{n1,n2,n3,n4,n5}. In this paper, we start testing the Kerr metric using the continuum-fitting method. We employ our new model {\sc nkbb} (Non-Kerr multi-temperature BlackBody spectrum)~\citep{nkbb} and we analyze 17~\textsl{RXTE} observations of the stellar-mass black hole in LMC~X-1 to constrain possible deviations from the Kerr solution.

\vspace{0.5cm}


\section{Testing black holes using the continuum-fitting method}

The continuum-fitting method is the analysis of the thermal spectrum of a geometrically thin and optically thick accretion disk around a black hole~\citep{cfm1}. Geometrically thin accretion disks are normally described by the Novikov-Thorne model~\citep{ntm1,ntm2}, which is the relativistic generalization of the Shakura-Sunyaev model~\citep{ss73}. The disk is assumed to be on the equatorial plane, perpendicular to the black hole spin, and the inner edge of the disk is set at the innermost stable circular orbit (ISCO). From the conservation of mass, energy, and angular momentum, we infer the time averaged radial structure of the disk. In the Kerr spacetime, the thermal spectrum of the disk turns out to depend only on five parameters: black hole distance $D$, inclination angle of the disk with respect to the line of sight of the observer $i$, black hole mass $M$, black hole spin parameter $a_*$, and mass accretion rate $\dot{M}$. Since the spectrum is degenerate with respect to these five parameters, spin measurements require independent estimates of $D$, $i$, and $M$, and the fit can provide the spin parameter $a_*$ and the mass accretion rate $\dot{M}$~\citep{cfm1,cfm2}.

{\sc nkbb} follows the so called bottom-up approach~\citep{nkbb}. The thermal spectrum of the accretion disk is calculated in a parametric black hole spacetime, where possible deviations from the Kerr geometry are quantified by introducing {\it ad hoc} deformation parameters. The spirit behind this approach is to perform a null-experiment and check whether astronomical data require that all deformation parameters vanish; that is, the Kerr metric is sufficient to model the data well. If a possible deviation from the Kerr solution is found, this approach is not suitable to measure the deviation from the Kerr geometry and it is necessary to use a different method. Among the many proposed parametric black hole spacetimes in literature, here we use the Johannsen metric~\citep{jj}.

In its simplest form, which is the version employed in our work, the Johannsen metric has only one deformation parameter, named $\alpha_{13}$ [see \citet{jj} for the origin of this parameter]. In Boyer-Lindquist coordinates, the line element reads (we use units in which $G_{\rm N} = c = 1$)
\be
ds^2 &=& - \frac{\Sigma \left(\Sigma - 2 M r \right)}{A^2} \, dt^2
+ \frac{\Sigma}{\Delta} \, dr^2 + \Sigma \, d\theta^2
+ \frac{\left[ \left(r^2 + a^2\right)^2 \left(1 + \delta\right)^2 
- a^2 \Delta \sin^2\theta\right] 
\Sigma \sin^2\theta}{A^2} \, d\phi^2
\nonumber\\ && 
- \frac{2 a \left[ 2 M r + \delta \left(r^2 + a^2\right) \right] 
\Sigma \sin^2\theta}{A^2} \, dt \, d\phi \, ,
\ee
where $a = a_* M$, $\Sigma = r^2 + a^2 \cos^2\theta$, $\Delta = r^2 - 2 M r + a^2$, and
\be
A = \Sigma + \delta \left(r^2 + a^2\right) \, , \quad
\delta = \alpha_{13} \left(\frac{M}{r}\right)^3 \, .
\ee
For $\alpha_{13} = 0$, we recover the Kerr metric. In order to have a regular exterior region, we have to impose $| a_* | \le 1$ (as in the Kerr metric, for $| a_* | > 1$ there is no event horizon and the central singularity is naked) and the following restriction on $\alpha_{13}$~\citep{ashu}
\be\label{eq-bound}
\alpha_{13} > - \frac{1}{2} \left( 1 + \sqrt{1 - a^2_*} \right)^4 \, .
\ee

Note that $\alpha_{13}$ enters the metric coefficients $g_{tt}$, $g_{t\phi}$, and $g_{\phi\phi}$. It thus affects the structure of the accretion disk, modifying the Keplerian gas motion and moving the ISCO radius. Qualitatively speaking, $\alpha_{13} > 0$ ($< 0$) increases (decreases) the strength of the gravitational force, so it moves the ISCO radius to higher (lower) values and this explains the strong correlation with the spin parameter in the plots that will be presented in the next sections.  

\vspace{0.5cm}


\section{Data reduction and analysis}

LMC~X-1 was discovered in 1969 as the first extragalactic X-ray binary~\citep{p1,p2}. The system consists of a stellar-mass black hole and an O-giant companion star. The distance of the source, the black hole mass, and the inclination angle of the orbit, which are three key-quantities in the continuum-fitting method, have been estimated to be $D = 48.10 \pm 2.22$~kpc, $M = 10.91 \pm 1.54$~$M_\odot$, and $i = 36.38 \pm 2.02$~deg, respectively~\citep{jo,lg}. LMC~X-1 is characterized by a quite stable bolometric luminosity, which is about 16\% of its Eddington luminosity $L_{\rm Edd}$~\citep{lg} and thus nicely meets the standard criterion required to use the continuum-fitting method: sources with an accretion luminosity in the range 5\% to 30\% $L_{\rm Edd}$~\citep{cfm2}.

The measurement of the spin parameter of the black hole in LMC~X-1 using the continuum-fitting method was presented in~\citet{lg}, and here we follow that study. There are 55 pointed observations of LMC~X-1 with the Proportional Counter Array (PCA) onboard \textsl{RXTE}~\citep{pca}. To use the continuum-fitting method, it is desirable to choose thermal dominant spectral data, which are defined by three conditions~\citep{rem}: $i)$ the flux of the thermal component accounts for more than 75\% of the total 2-20~keV unabsorbed flux, $ii)$ the root mean square (RMS) variability in the power density spectrum in the 0.1-10~Hz range is lower than 0.075, and $iii)$ quasi-periodic oscillations (QPOs) are absent or very weak. Imposing these three conditions, we only have 3~observations, which are named ``gold spectra'' in~\citet{lg} as they are supposed to be more suitable for the continuum-fitting method. Relaxing condition~$i)$, we have 14~more observations, which are named ``silver spectra'' in~\citet{lg}\footnote{We note that in~\citet{lg} there are 15~silver spectra, but here we ignore one of these observations because of a problem in the current version of the ftool {\sc rbnrmf}.}. Additionally, we require that the accretion luminosity is in the range 5\% to 30\% in order to assure that the accretion disk is geometrically thin and the inner edge is at the ISCO~\citep{cfm2}, but this is always satisfied in the \textsl{RXTE} observations of LMC~X-1.

In our study, we analyzed 17~observations of the Proportional Counter Array (PCA). PCA was on board of the \textsl{Rossi X-ray Timing Explorer} (\textsl{RXTE}), which was launched in 1995 and decommissioned in 2012. PCA was designed to study far-away faint sources in the 2-60~keV energy range and consisted of 5 proportional counter units (PCUs), which comprised of Xenon layers to detect photons. We used the Heasoft version~6.25 to reduce the data and unprocessed data files were downloaded from the HEASARC website.

We used the pulse-height spectra of only PCU-2 because it is the best calibrated PCU and is the most operational one. ``Standard 2'' mode data were used for reduction. To derive the spectra, data from all the Xenon layers were combined for PCU-2. Background spectra were obtained by the latest ``faint source'' background model provided by the \textsl{RXTE} team and using the ftool {\sc pcabackest}. The final spectra were then obtained by subtracting the background spectra from the total spectra. The response files were constructed and combined for each layer using the ftool {\sc pcarsp}. The data were corrected for calibration using the python script {\sc pcacorr}~\citep{Garcia:2014cra}. Finally, we added a systematic error of 0.1\% to all the PCA energy channels.

We only used data in the energy range 3-20~keV. Below 3~keV, calibration effects become dominant [see, e.g., \citet{Jahoda:2005uw}] and, in particular, the {\sc pcacorr} correction is not valid in this regime. Above 20 keV, the background becomes dominant and the calibration is uncertain.

We fit each of the 17~observations with the XSPEC model~\citep{xspec}

\vspace{0.2cm}

{\sc TBabs$\times$(simpl$\times$nkbb)} .

\vspace{0.2cm}

\noindent {\sc TBabs} describes the Galactic absorption~\citep{tbabs}. We use the abundances of \citet{tbabs} and we freeze the hydrogen column density to $N_{\rm H} = 4.6 \cdot 10^{21}$~cm$^{-2}$~\citep{jo}; however, its exact value does not matter considering it is low and we are analyzing \textsl{RXTE} data that do no cover the low energies that are especially sensitive to low column densities. {\sc nkbb} describes the thermal spectrum of the accretion disk~\citep{nkbb}. The distance of the source $D$, the black hole mass $M$, and the inclination angle of the orbit $i$ are frozen to 48.10~kpc, 10.91~$M_\odot$, and 36.38~deg, respectively; at this stage, we ignore their uncertainties. The hardening factor is frozen to 1.55 for all observations, which is the value found in~\citet{lg}. We checked that its impact is weak and we get very similar results even if we use 1.45 or 1.65. The hardening factor varies with luminosity, but LMC~X-1 varies modestly around 16\% of its Eddington limit\footnote{In the \textsl{RXTE} observations analyzed in this work, $L/L_{\rm Edd}$ is in the range 0.145 to 0.171, see Tab.~2 in \citet{lg}.}, which justifies the use of a fixed value. The spin parameter $a_*$ and the mass accretion rate are always free parameters to be determined by the fit. The deformation parameter $\alpha_{13}$ is first set to zero (Kerr metric) in order to check if we can recover the results of~\citet{lg}, and then it is left free in order to measure possible deviations from the Kerr spacetime. {\sc simpl} converts a fraction $f_{\rm SC}$ of thermal photons into a power-law-like spectrum with photon index $\Gamma$ to describe the radiation from the corona~\citep{simpl}, providing a superior description of the Comptonization at low energies as compared to a power-law. Since the data do not permit us to determine $\Gamma$, we freeze it to 2.5, as in~\citet{lg}. Employing a different value for $\Gamma$ has a marginal impact on the estimate of the other parameters~\citep{lg}. In the end, the model has 3~free parameters ($a_*$, $\dot{M}$, and $f_{\rm SC}$) when we assume the Kerr metric and 4 (even $\alpha_{13}$) otherwise.

\begin{table*}
\centering
\caption{ \label{t-fit}}
{\renewcommand{\arraystretch}{1.3}
\begin{tabular}{lc|cccc|ccccc}
\hline\hline
No. & UT & \multicolumn{4}{c}{$\alpha_{13} = 0$} & \multicolumn{5}{c}{$\alpha_{13}$ free} \\
&& $a_*$ & $\dot{M}$ & $f_{\rm SC}$ & $\chi^2$/dof & $a_*$ & $\alpha_{13}$ & $\dot{M}$ & $f_{\rm SC}$ & $\chi^2$/dof \\
\hline
1 & 1996-06-09 & $0.936_{-0.015}^{+0.014}$ & $1.39_{-0.08}^{+0.09}$ & $0.073_{-0.005}^{+0.005}$ & 38.08/42 & $0.992_{-0.30}^{\rm +(P)}$ & $0.29_{-2.2}^{+0.05}$ & $1.34_{-0.05}^{+0.15}$ & $0.073_{-0.005}^{+0.005}$ & 37.99/41 \\
2 & 1996-08-01 & $0.881_{-0.018}^{+0.017}$ & $1.86_{-0.10}^{+0.10}$ & $0.066_{-0.004}^{+0.004}$ & 23.13/42 & $0.998_{-0.20}$ & $0.59_{\rm -(P)}^{+0.10}$ & $1.91_{-0.23}^{+0.03}$ & $0.066_{-0.004}^{+0.004}$ & 23.08/41 \\
3 & 1997-03-09 & $0.943_{-0.010}^{+0.009}$ & $1.47_{-0.07}^{+0.07}$ & $0.050_{-0.004}^{+0.004}$ & 45.15/42 & $0.998_{-0.27}$ & $0.29_{-1.7}^{+0.01}$ & $1.44_{-0.10}^{+0.07}$ & $0.050_{-0.004}^{+0.004}$ & 45.02/41 \\
4 & 1997-03-21 & $0.948_{-0.008}^{+0.007}$ & $1.48_{-0.05}^{+0.05}$ & $0.046_{-0.003}^{+0.003}$ & 34.48/42 & $0.998_{-0.28}$ & $0.27_{-2.0}^{+0.04}$ & $1.45_{-0.02}^{+0.22}$ & $0.046_{-0.003}^{+0.003}$ & 34.24/41 \\
5 & 1997-04-16 & $0.921_{-0.010}^{+0.009}$ & $1.64_{-0.06}^{+0.06}$ & $0.056_{-0.003}^{+0.003}$ & 45.47/42 & $0.998_{-0.24}$ & $0.39_{-1.6}^{+0.01}$ & $1.68_{-0.16}^{+0.23}$ & $0.056_{-0.003}^{+0.003}$ & 45.07/41 \\
\hline
6 & 1997-05-07 & $0.938_{-0.008}^{+0.008}$ & $1.55_{-0.05}^{+0.06}$ & $0.053_{-0.003}^{+0.003}$ & 37.71/42 & $0.993_{-0.48}^{\rm +(P)}$ & $0.29_{-2.2}^{+0.05}$ & $1.50_{-0.04}^{+0.12}$ & $0.053_{-0.003}^{+0.003}$ & 37.51/41 \\
7 & 1997-05-28 & $0.964_{-0.015}^{+0.012}$ & $1.31_{-0.10}^{+0.11}$ & $0.058_{-0.007}^{+0.007}$ & 29.01/42 & $0.995_{-0.34}^{\rm +(P)}$ & $0.2_{-3.1}^{+0.2}$ & $1.28_{-0.18}^{+0.16}$ & $0.058_{-0.009}^{+0.007}$ & 28.92/41 \\
8 & 1997-05-29 & $0.938_{-0.018}^{+0.016}$ & $1.42_{-0.10}^{+0.11}$ & $0.049_{-0.006}^{+0.006}$ & 29.83/42 & $0.996_{-0.40}^{\rm +(P)}$ & $0.31_{-4.4}^{+0.01}$ & $1.38_{-0.05}^{+0.05}$ & $0.049_{-0.005}^{+0.006}$ & 29.76/41 \\
9 & 1997-07-09 & $0.917_{-0.019}^{+0.017}$ & $1.57_{-0.10}^{+0.11}$ & $0.046_{-0.005}^{+0.005}$ & 28.33/42 & $0.274_{-0.075}^{+0.003}$ & $-5^{+0.05}$ & $1.014_{-0.204}^{+0.005}$ & $0.049_{-0.005}^{+0.003}$ & 27.71/41 \\
10 & 1997-08-20 & $0.945_{-0.008}^{+0.007}$ & $1.57_{-0.06}^{+0.06}$ & $0.048_{-0.003}^{+0.003}$ & 34.47/42 & $0.998_{-0.34}$ & $0.29_{-2.9}^{+0.05}$ & $1.53_{-0.03}^{+0.04}$ & $0.048_{-0.003}^{+0.003}$ & 34.36/41 \\
\hline
11$^\star$ & 1997-09-12 & $0.906_{-0.011}^{+0.010}$ & $1.76_{-0.06}^{+0.06}$ & $0.048_{-0.003}^{+0.003}$ & 31.29/42 & $0.2234_{-0.0014}^{+0.0424}$ & $-5^{+0.05}$ & $1.189_{-0.085}^{+0.003}$ & $0.051_{-0.003}^{+0.002}$ & 29.62/41 \\
12 & 1997-09-19 & $0.964_{-0.010}^{+0.008}$ & $1.31_{-0.06}^{+0.07}$ & $0.063_{-0.004}^{+0.004}$ & 28.92/42 & $0.994_{-0.19}^{\rm +(P)}$ & $0.18_{-0.72}^{+0.02}$ & $1.27_{-0.13}^{+0.10}$ & $0.063_{-0.004}^{+0.004}$ & 28.77/41 \\
13 & 1997-12-12 & $0.925_{-0.011}^{+0.010}$ & $1.65_{-0.06}^{+0.07}$ & $0.042_{-0.003}^{+0.003}$ & 26.01/42 & $0.93_{-0.82}^{\rm +(P)}$ & $0.0_{\rm -(P)}^{+0.3}$ & $1.65_{-0.10}^{+0.08}$ & $0.042_{-0.003}^{+0.003}$ & 26.01/41 \\
14 & 1998-03-12 & $0.966_{-0.007}^{+0.007}$ & $1.40_{-0.04}^{+0.06}$ & $0.054_{-0.004}^{+0.004}$ & 30.37/42 & $0.998_{-0.20}$ & $0.19_{-0.44}^{+0.21}$ & $1.39_{-0.17}^{+0.05}$ & $0.054_{-0.004}^{+0.004}$ & 30.36/41 \\
15 & 1998-05-06 & $0.948_{-0.009}^{+0.008}$ & $1.43_{-0.06}^{+0.06}$ & $0.048_{-0.003}^{+0.003}$ & 28.01/42 & $0.998_{-0.43}$ & $0.27_{-2.5}^{+0.05}$ & $1.40_{-0.04}^{+0.12}$ & $0.048_{-0.003}^{+0.003}$ & 27.90/41 \\
\hline
16$^\star$ & 1998-07-20 & $0.936_{-0.009}^{+0.009}$ & $1.52_{-0.05}^{+0.06}$ & $0.042_{-0.003}^{+0.003}$ & 21.07/42 & $0.55_{-0.20}^{+0.28}$ & $-2.9_{\rm -(P)}^{+3.2}$ & $1.19_{-0.36}^{+0.15}$ & $0.044_{-0.003}^{+0.003}$ & 20.50/41 \\
17$^\star$ & 2004-01-07 & $0.935_{-0.022}^{+0.018}$ & $1.45_{-0.13}^{+0.13}$ & $0.046_{-0.005}^{+0.005}$ & 19.02/42 & $0.54_{-0.23}^{+0.26}$ & $-3.0_{\rm -(P)}^{+3.3}$ & $1.1_{-0.4}^{+0.3}$ & $0.048_{-0.005}^{+0.005}$ & 18.96/41 \\
\hline\hline
\end{tabular}}
\tablenotetext{0}{Best-fit values for observations~1-17 assuming the Kerr spacetime ($\alpha_{13} = 0$) and without such an assumption ($\alpha_{13}$ free). The reported uncertainties correspond to the 90\% confidence level for one relevant parameter ($\Delta\chi^2 = 2.71$). Note that the maximum value of $a_*$ allowed by the model is 0.998, while the minimum value of $\alpha_{13}$ is $-5$. In several cases, the best fit is stuck at $a_* = 0.998$ or at $\alpha_{13} = -5$ and therefore we do not report the upper/lower uncertainty. (P) means that the 90\% confidence level reaches the maximum/minimum value of the parameter. $^\star$ marks the three ``golden spectra'', which meet all the conditions for thermal dominant spectral data.} 
\vspace{0.2cm}
\end{table*}

\vspace{0.5cm}
  
\section{Results}

\begin{figure*}[t]
\vspace{-1.5cm}
\begin{center}
\includegraphics[type=pdf,ext=.pdf,read=.pdf,width=19.0cm]{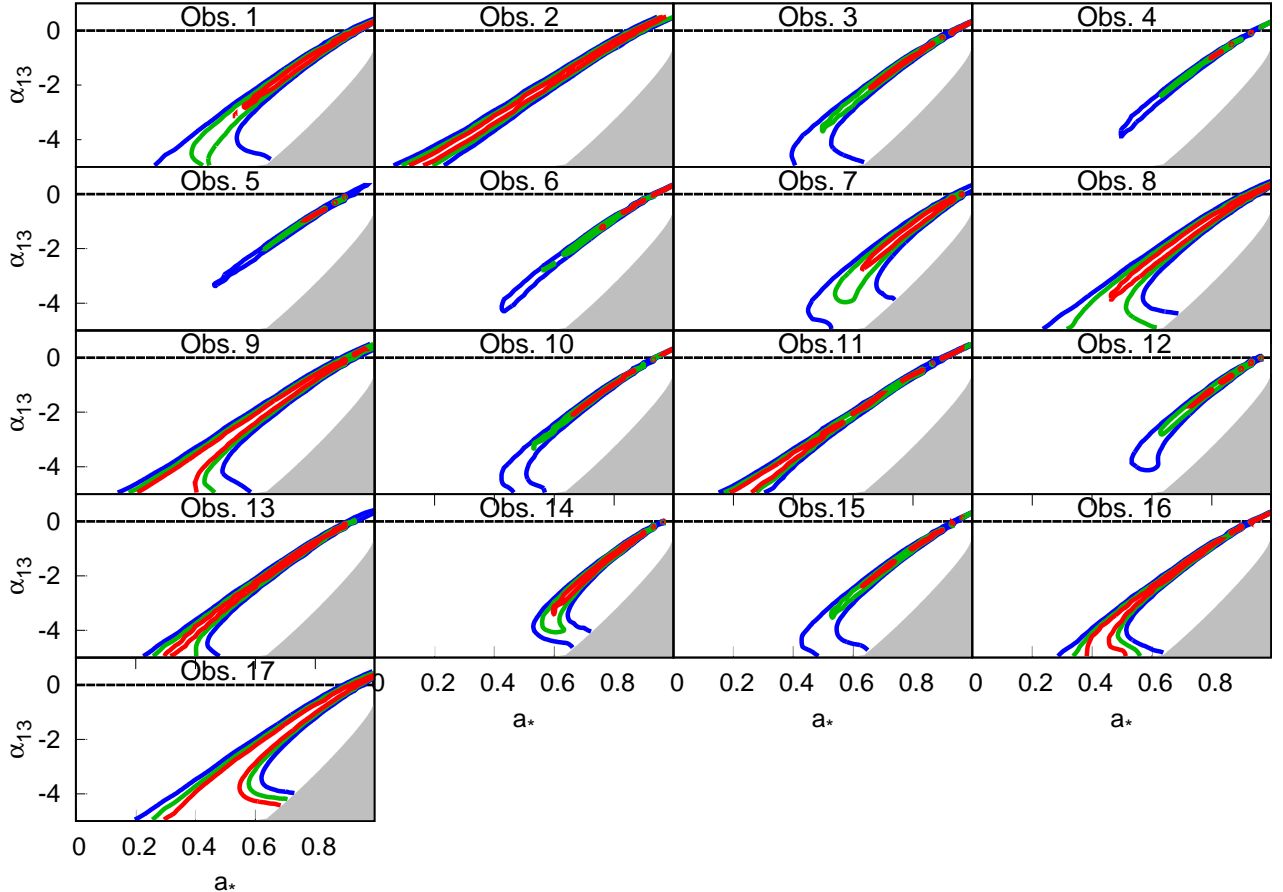}
\end{center}
\vspace{-1.5cm}
\caption{Constraints on the spin parameter $a_*$ and the deformation parameter $\alpha_{13}$ for observations~1-17. The red, green, and blue curves represent, respectively, the 68\%, 90\% and 99\% confidence level curves for two relevant parameters ($\Delta\chi^2 = 2.30$, 4.61, and 9.21, respectively). The gray region is ignored because includes pathological spacetimes; see Eq.~(\ref{eq-bound}). \label{f-cplots}}
\vspace{0.5cm}
\end{figure*}

The best-fit values of our 17~observations are reported in Tab.~\ref{t-fit} for $\alpha_{13} = 0$ (left column) and for $\alpha_{13}$ free (right column). The constraints on the spin parameter $a_*$ and the deformation parameter $\alpha_{13}$ for every observation are shown in Fig.~\ref{f-cplots}. The analysis is done with the standard XSPEC routines and the plots in Fig.~\ref{f-cplots} are calculated with the {\sc steppar} command in XSPEC.

Roughly speaking, the continuum fitting method measures the position of the inner edge of the disk, which is set at the ISCO in the Novikov-Thorne model and only depends on the spin parameter in the Kerr spacetime. This allows us to measure the black hole spin when we assume the Kerr metric. Relaxing the Kerr hypothesis, the situation is more complicated. As we can see from Fig.~\ref{f-cplots}, there is a strong correlation between $a_*$ and $\alpha_{13}$. Now the ISCO is set by $a_*$ and $\alpha_{13}$ [see~\citet{kong} for more details], so it is difficult to constrain the values of $a_*$ and $\alpha_{13}$.

\begin{figure}[t]
\begin{center}
\includegraphics[type=pdf,ext=.pdf,read=.pdf,width=8.5cm]{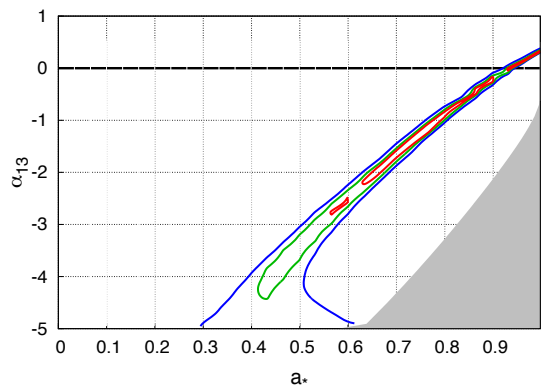}
\end{center}
\vspace{-1.3cm}
\caption{As in Fig.~\ref{f-cplots} when we combine all observations together. See the text for more details. \label{f-all}}
\end{figure}

We combine all observations together following the standard approach of averaging the $\chi^2$ at each grid-point in the $(a_*,\alpha_{13})$ plane. The measurement of $a_*$ and $\alpha_{13}$ is (90\% confidence level for one relevant parameter)
\be\label{eq-final-mes}
a_* = 0.998_{-0.44} \, , \quad \alpha_{13} = 0.32_{-3.1}^{+0.04} \, .
\ee
Fig.~\ref{f-all} shows the constraints on the spin parameter $a_*$ and the deformation parameter $\alpha_{13}$ at the 68\%, 90\% and 99\% of confidence level for two relevant parameters ($\Delta\chi^2 = 2.30$, 4.61, and 9.21, respectively).

\begin{figure*}
\begin{center}
\includegraphics[type=pdf,ext=.pdf,read=.pdf,width=8.5cm,trim={1.0cm 0.5cm 4.0cm 13.0cm},clip]{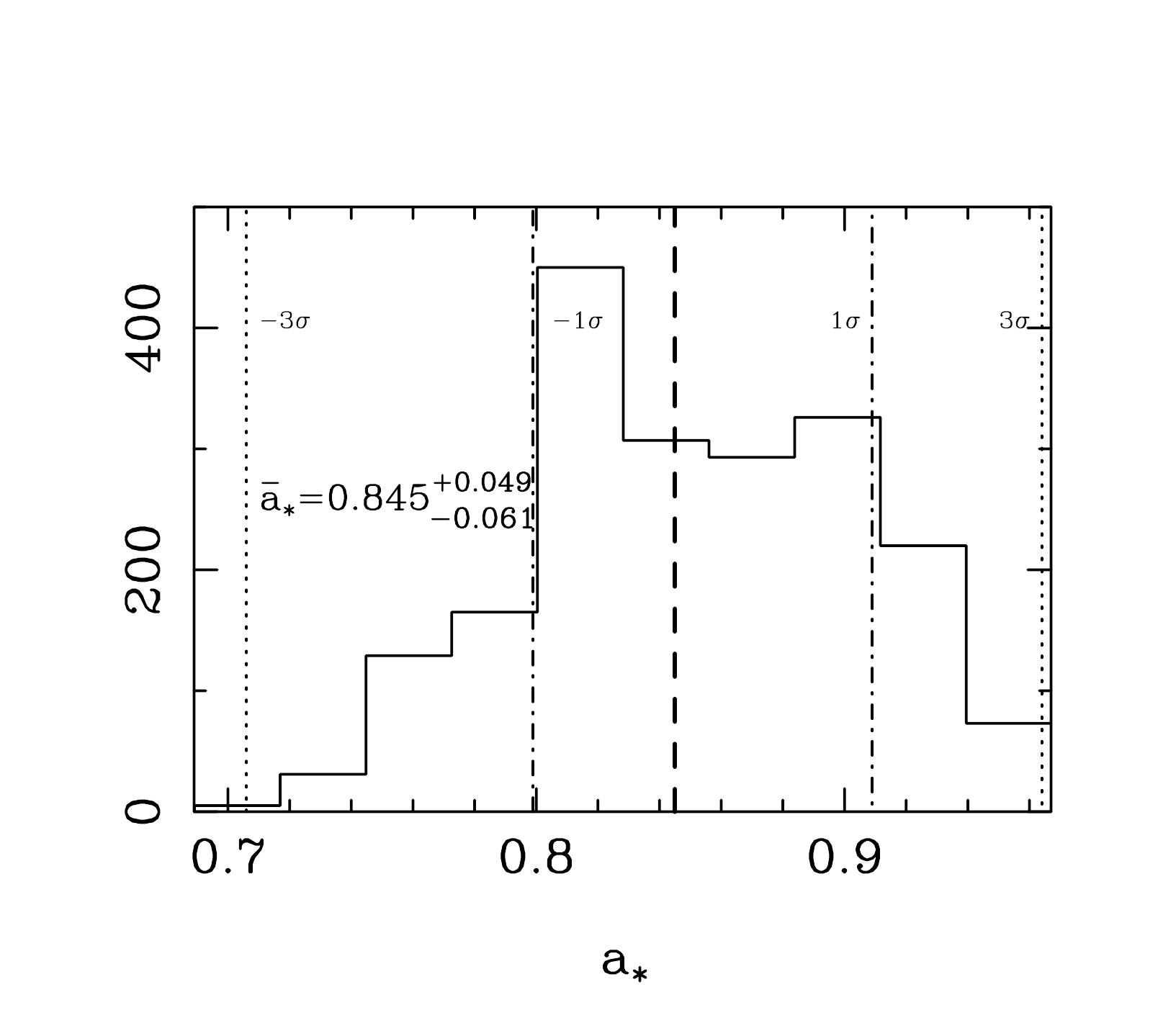}
\hspace{0.5cm}
\includegraphics[type=pdf,ext=.pdf,read=.pdf,width=8.5cm,trim={1.0cm 0.5cm 4.0cm 13.0cm},clip]{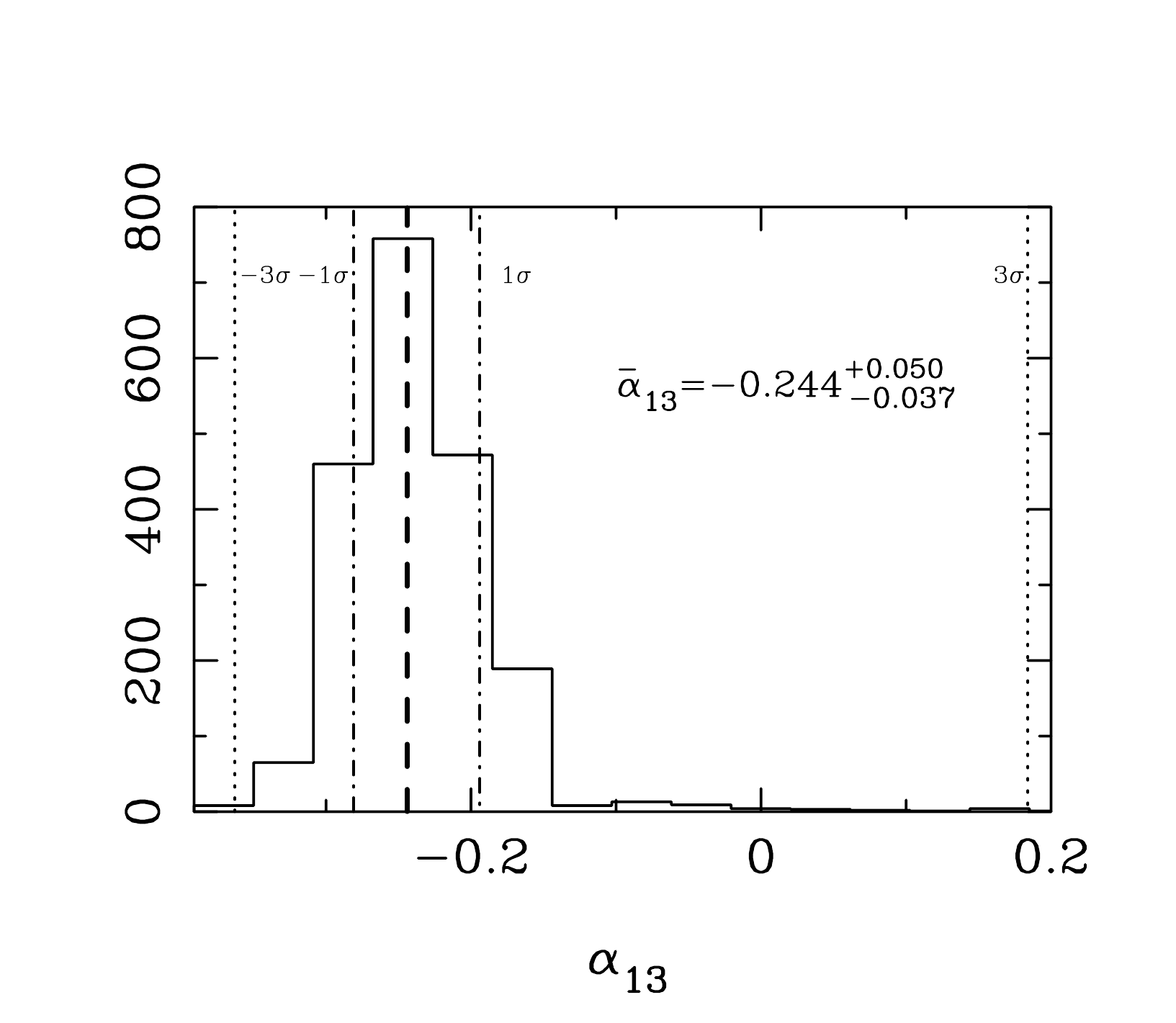}
\end{center}
\vspace{-0.7cm}
\caption{Histograms of the spin parameter $a_*$ (left panel) and of the deformation parameter $\alpha_{13}$ (right panel) for 2,000 sets of parameters $(M,D,i)$. The thick-dashed vertical lines mark the mean value of the fitting parameter, while the thin-dotted-dashed and thin-dotted vertical lines mark, respectively, the 1-$\sigma$ and 3-$\sigma$ limits. See the text for more details. \label{f-histo}}
\end{figure*}

In the Kerr spacetime, the uncertainty on the final measurement of the black hole spin parameter is dominated by the observational uncertainties on $D$, $M$, and $i$. In order to evaluate the impact of the observational uncertainties of these three parameters on our measurements of $a_*$ and $\alpha_{13}$, we proceed as in~\citet{lg}. We generate 2,000 parameter sets for $D$, $M$, and $i$ with Monte Carlo simulations assuming that the uncertainties in these parameters are normally and independently distributed. For every set, we find the best-fit of the combined observations. We thus calculate the median values of the best-fit values of the spin parameter $a_*$ and of the deformation parameter $\alpha_{13}$. Our result is shown in Fig.~\ref{f-histo}. Contrary to the Kerr case, in which the uncertainties of $D$, $M$, and $i$ provide the main contribution on the final uncertainty on $a_*$, in the non-Kerr case their impact is small and eventually subdominant with respect to the statistical uncertainty shown in Eq.~(\ref{eq-final-mes}). This is because the degeneracy between $a_*$ and $\alpha_{13}$ plays an important role in the final measurement. 
Neglecting the subdominant contributions from the uncertainties of $D$, $M$, and $i$, our final measurement of $a_*$ and $\alpha_{13}$ is thus given in Eq.~(\ref{eq-final-mes}).

\vspace{0.5cm}
  
\section{Concluding remarks}

In this paper, we have presented the first attempt to use the continuum-fitting method with real data and test the Kerr nature of an astrophysical black hole.
We have employed the new model {\sc nkbb} to fit 17~\textsl{RXTE} observations of the X-ray binary LMC~X-1 and constrain the deformation parameter $\alpha_{13}$ of the Johannsen metric. In the past couple of years, similar tests of the Kerr metric using \textsl{XMM-Newton}, \textsl{Suzaku}, and \textsl{NuSTAR} data were done by analyzing the reflection spectrum, never with the thermal spectrum.

We find that it is quite challenging to constrain $a_*$ and $\alpha_{13}$ at the same time. This point is clear if we compare the spin measurement reported in~\citet{lg} assuming the Kerr metric, $a_* = 0.92_{-0.07}^{+0.05}$, and our spin measurement with free $\alpha_{13}$, $a_* > 0.56$. When $\alpha_{13}$ is free, the constraint on $a_*$ is much weaker. Moreover, in the traditional continuum-fitting method for the Kerr metric, the main sources of uncertainty in the final spin measurement are the uncertainties on the black hole mass, distance, and inclination angle of the disk~\citep{2011MNRAS.414.1183K,cfm2}, three quantities that must be determined from independent measurements, often with optical observations. When we test the Kerr metric, the uncertainties on these three parameters seem to be subdominant. Even assuming they are known without uncertainty, the intrinsic degeneracy between the black hole spin and the deformation parameter does not permit precise measurements of $a_*$ and $\alpha_{13}$.

The problem of testing the Kerr metric with electromagnetic data without an independent measurement of the black hole spin parameter is well known in literature~\citep[see, e.g.,][]{2012ApJ...754..133K,2013ApJ...773...57J,kong,2016PhRvD..93d4020H}. We usually meet this issue when the deformation parameter affects the metric coefficients $g_{tt}$, $g_{t\phi}$, and/or $g_{\phi\phi}$, which can have a strong impact on the location of the ISCO radius and, in turn, on the shape of the spectrum. This is the case of the deformation parameter $\alpha_{13}$ of the Johannsen metric and the simple shape of the thermal spectrum of the disk, which is simply a multi-temperature blackbody spectrum, cannot break the parameter degeneracy. Note that the correlation between the measurements of $a_*$ and $\alpha_{13}$ is usually quite similar among different electromagnetic techniques, since all of them are mainly sensitive to the exact location of the inner edge of the disk, while other relativistic effects more specific of the particular spectral component have often a weaker impact.

A comparison between the constraints in Fig.~\ref{f-all} and those found from the analysis of the reflection spectrum of the disk of other sources in previous studies is not straightforward because these measurements are quite sensitive to the specific source and the quality of the data, so it may be dangerous to generalize the results found from our analysis of 17~\textsl{RXTE} observations of LMC~X-1. 
In general, a correlation between the estimates of $a_*$ and $\alpha_{13}$ is common even when we analyze the reflection spectrum. However, such a degeneracy can be broken when the inner edge of the disk is very close to the compact object~\citep{ashu,n3,n5}. For example, \citet{n2} analyzed simultaneous \textsl{XMM-Newton} and \textsl{NuSTAR} observations of MCG--6--30--15 obtaining $a_* = 0.976_{-0.013}^{+0.007}$ and $\alpha_{13} = 0.00_{-0.20}^{+0.07}$ (90\% confidence level for one relevant parameter). Even if there is a correlation between these measurements of $a_*$ and $\alpha_{13}$, see Fig.~6 in \citet{n2}, we can get quite stringent constraints on both parameters. This is not the case with the analysis of the thermal component presented in this paper. When we assume the Kerr metric, observations~7, 12, and 14 of LMC~X-1 give quite high spin values (see left column of Tab.~\ref{t-fit}), comparable to the \textsl{XMM-Newton} and \textsl{NuSTAR} observations of MCG--6--30--15. However, when we leave $\alpha_{13}$ free it is not easy to constrain $a_*$ and $\alpha_{13}$ at the same time.
While the limited energy resolution of \textsl{RXTE} with respect to \textsl{XMM-Newton} may have some effect, the key-point is in the difference between the reflection spectrum and the thermal one. The former is characterized by many features, notably, but not only, the iron K$\alpha$ complex around 6-7~keV. Such features help to break the parameter degeneracy, even if the reflection spectrum has several parameters to fit. The thermal spectrum, on the contrary, has quite a simple shape and there is an intrinsic degeneracy among the model parameters. This is true even when we assume the Kerr metric: it is possible to measure the black hole spin only when we have independent estimates of the black hole mass, distance, and inclination angle of the disk. If we want to use the continuum-fitting method to test the Kerr metric and we add a deformation parameter, the problem of degeneracy between $a_*$ and $\alpha_{13}$ should not surprise.

While the analysis of the reflection spectrum of an accreting black hole is likely a more powerful technique for getting stringent constraints on possible deviations from the Kerr geometry, we can expect that the combination of the two methods to test the same source can provide more reliable and stronger constraints. In general, the possibility of a combined analysis is not automatic, because the two methods require, respectively, a strong reflection component and a strong thermal component. Some sources do not have data suitable for both techniques. Moreover, the continuum-fitting method requires independent estimates of the black hole mass, distance, and inclination angle of the disk, while most sources do not have reliable measurements of these three quantities. We plan to present the combined constraints from the analysis of the reflection and thermal components of the same source in a forthcoming paper.


\vspace{0.3cm}

This work was supported by the Innovation Program of the Shanghai Municipal Education Commission, Grant No.~2019-01-07-00-07-E00035, and the National Natural Science Foundation of China (NSFC), Grant No.~11973019.
V.G. is supported through the Margarete von Wrangell fellowship by the ESF and the Ministry of Science, Research and the Arts Baden-W\"urttemberg.
A.T., C.B., V.G., H.L., and J.F.S. are members of the International Team~458 at the International Space Science Institute (ISSI), Bern, Switzerland, and acknowledge support from ISSI during the meetings in Bern.

\vspace{0.3cm}


\end{document}